\documentclass[useAMS,usenatbib]{mn2e}

\usepackage{times}
\usepackage{amsmath}
\usepackage{amssymb}
\usepackage{graphicx}
\usepackage{graphics}

\title[Large Scale Clumpy Structure at $z\sim1.27$]
{Discovery of a Large Scale Clumpy Structure around the Lynx Supercluster
at $z\sim1.27$\thanks{Based on data collected at Subaru Telescope, which is
operated by the National Astronomical Observatory of Japan.}}
\author[Nakata et al.]{Fumiaki
Nakata,$^{1,2,3}$\thanks{E-mail: fumiaki.nakata@durham.ac.uk}
Tadayuki Kodama,$^{2,3}$ Kazuhiro Shimasaku,$^{3}$ Mamoru Doi,$^{4}$
\newauthor
Hisanori Furusawa,$^{5}$ Masaru Hamabe,$^{6}$ Masahiko Kimura,$^{7}$
Yutaka Komiyama,$^{5}$
\newauthor
Satoshi Miyazaki,$^{5}$ Sadanori Okamura,$^{3}$
Masami Ouchi,$^{8}$\thanks{Hubble Fellow}
Maki Sekiguchi,$^{9}$
\newauthor
Yoshihiro Ueda$^{10}$, Masafumi Yagi$^{2}$ and Naoki Yasuda$^{9}$
\\
$^{1}$Department of Physics, University of Durham, South Road, Durham
DH1 3LE, UK \\
$^{2}$National Astronomical Observatory of Japan, Mitaka, Tokyo
181-8588, Japan \\
$^{3}$Department of Astronomy, School of Science, University of Tokyo,
Bunkyo-ku, Tokyo 113-0033, Japan \\
$^{4}$Institute of Astronomy, University of Tokyo, Mitaka, Tokyo
181-1500, Japan \\
$^{5}$Subaru Telescope, National Astronomical Observatory of Japan,
Hilo, HI 96720, U.S.A \\
$^{6}$Department of Mathematical and Physical Sciences, Faculty of
Science, Japan Women's University, Bunkyo-ku, Tokyo 112-8681, Japan \\
$^{7}$Department of Astronomy, Faculty of Science, Kyoto University,
Sakyo-ku, Kyoto 606-8502, Japan \\
$^{8}$Space Telescope Science Institute, 3700 San Martin Drive,
Baltimore, MD 21210, U.S.A. \\
$^{9}$Institute for Cosmic Ray Research, University of Tokyo, Kashiwa,
Chiba 277-8582, Japan \\
$^{10}$Institute of Space and Astronautical Science, 
3-1-1 Yoshinodai, Sagamihara, Kanagawa 229-8510, Japan \\
}

\begin{document}

\date{Accepted ???. Received ???; in original form ???}

\pagerange{\pageref{firstpage}--\pageref{lastpage}} \pubyear{2004}

\maketitle

\label{firstpage}

\begin{abstract}
We report the discovery of a probable large scale structure composed of many
galaxy clumps around the known twin clusters at $z=1.26$ and $z=1.27$
in the Lynx region. Our analysis is based on deep, panoramic, and multi-colour
imaging (26.4$'$$\times$24.1$'$) in $VRi'z'$-bands with the Suprime-Cam on
the 8.2~m Subaru telescope. This unique, deep and wide-field imaging data-set
allows us for the first time to map out the galaxy distribution in the highest
redshift supercluster known. We apply a photometric redshift technique to
extract plausible cluster members at $z\sim1.27$ down to $i'$=26.15 (5$\sigma$)
corresponding to $\sim M^\ast+2.5$ at this redshift.
From the 2-D distribution of these photometrically selected galaxies, we
newly identify seven candidates of galaxy groups or clusters where the surface
density of red galaxies is significantly high ($>$5$\sigma$),
in addition to the two known clusters. These candidates show clear red
colour-magnitude sequences consistent with a passive evolution model,
which suggests the existence of additional high density regions around the
Lynx superclusters.
\end{abstract}


\section{Introduction}

Superclusters are the largest systems of galaxies, composed of multiple
clusters of galaxies, extending over 10-20~Mpc (e.g., Bahcall,
Soneira 1984; Postman, Geller, Huchra 1988; Quintana et al.\ 1995; Small
et al.\ 1998).
According to the N-body simulations which successfully reproduce
the observed filamentary structures at the local Universe
(e.g., Peacock et al.\ 2001), clusters of galaxies at $z\sim1$ are
still in the process of formation (e.g., Moore et al.\ 1998).
Therefore, clusters of galaxies at this cosmological distance, especially
superclusters, are important sites where we can directly see the
process of structure formation and evolution and mass assembly
to cluster cores.

Despite the importance in the cosmological context, however, only a
few superclusters with the scale of $>$10~Mpc have been identified so
far at high redshifts, especially beyond $z\sim1$
(Connolly et al.\ 1996; Lubin et al.\ 2000; Tanaka et al.\ 2001).
The limitation comes largely from the small field of view of
large telescopes.

The advent of the Suprime-Cam (Miyazaki et al.\ 2002), a huge format
optical camera with a 30$'$ field of view on the prime focus of the 8.2m
Subaru telescope, has made it possible for us to view a $>$10~Mpc
region all at once at high redshifts.
Kodama et al.\ (2001) have taken this unique advantage of the Suprime-Cam
on Subaru to map out the large scale structure around the A851 cluster at
$z=0.41$. They have found many filamentary structures and subclumps spreading 
over 5~Mpc from the dense cluster core, on the basis of photometric
redshift (photo-z) analysis, an approach similar to that taken in
this {\it Letter}.

By utilizing the great light-collection power of the Subaru telescope,
we now target the most distant supercluster ever firmly identified with
spectroscopy: ie., the Lynx supercluster region at $z\sim1.27$ composed
of two known clusters, RXJ~0848.9+4452 at $z=1.26$ and RXJ~0848.6+4453 at
$z=1.27$, which are spectroscopically confirmed X-ray emitting
clusters firstly reported by Rosati et al.\ (1999) and Stanford et
al.\ (1997). The bolometric X-ray luminosities of RXJ~0848.9 and
RXJ~0848.6 are $0.69^{+0.27}_{-0.17}\times10^{44}$~ergs~s$^{-1}$
and $3.3^{+0.9}_{-0.5}\times10^{44}$~ergs~s$^{-1}$, respectively,
which are derived with the Chandra observation (Stanford et al.\ 2001).

Throughout this paper, we use the AB magnitude system (Oke, Gunn 1983).
The adopted cosmological parameters are $\Omega_0=0.3$ and $\lambda_0=0.7$,
which gives a physical scale of 8.37$h^{-1}_{70}$~kpc~arcsec$^{-1}$ at the
cluster redshift.

\section{Observations, Reduction, and Analysis}

\subsection{Observations and Reduction}

%
%

\begin{table}
\caption{Log of the observations.}
\label{table1}
\begin{tabular}{lcccc}
\hline
Band & Date & Total Exp. & lim. mag & Seeing \\
 & & & ($5\sigma$) & \\
\hline
  $V$  & 2000 Nov. 27 & 96 min & 26.78 & $\sim0.$\hspace{-2pt}$''$9 \\
  $R$  & 2000 Nov. 24 & 90 min & 26.63 & $\sim0.$\hspace{-2pt}$''$9 \\
  $i'$ & 2000 Nov. 25 & 30 min & 26.15 & $\sim0.$\hspace{-2pt}$''$8 \\
       & 2001 Mar. 21 & 30 min &       & $\sim0.$\hspace{-2pt}$''$8 \\
  $z'$ & 2001 Mar. 22 & 54 min & 24.88 & $\sim1.$\hspace{-2pt}$''$0 \\
\hline
\end{tabular}
\end{table}

The Lynx field was observed in November 2000--March 2001
in the $V$, $R$, $i'$, and $z'$ bands with the Suprime-Cam.
Only eight of the ten CCDs of the Suprime-Cam were ready at the time
of these observations. Consequently, each image has an FOV of
$27'\times 27'$ with a resolution of 0.\hspace{-2pt}$''$2 per pixel.
The log of the observations is given in Table~\ref{table1}.

The images are processed in a standard manner
with {\sc iraf} and purpose-written software
developed by us (Yagi et al.\ 2002).
Flux calibration is performed with photometric standard stars 
from Landolt (1992) for $V$- and $R$-band, and spectrophotometric standard
stars (Oke 1990; Bohlin, Colina, Finley 1995; Bohlin 1996) for $
i'$- and $z'$-band.
We register the $V,$ $R,$ and $i'$ images with the $z'$ image, and
match their PSFs to a fixed value, $1''$ (FWHM). Taking an overlapped
region for all the images in the four bands, the final area used in
our analysis is restricted in a $26'.4\times24'.1$ region, which
corresponds to  $13.3\times12.1h^{-2}_{70}$~Mpc$^2$ at $z=1.27$.
The regions at the edges of the combined image and
regions around bright stars
were masked and not used in the analysis.

Since the $i'$-band image is the deepest among the four bands with respect to
the passively evolving galaxies at the cluster redshift ($i'$(5$\sigma$)=26.15
mag or $\sim M^\ast+2.5$ at $z=1.27$), we constructed an $i'$-band selected
sample using SExtractor v.2.2.0 (Bertin, Arnouts 1996). An area with more
than 5 connected pixels with counts over 2.8$\sigma_{\rm sky}$ is identified
as an object. Photometry was then performed with a relatively small diameter
aperture of 2$''$ (16.7$h^{-1}_{70}$~kpc) to keep the signal-to-noise ratios
high enough for photo-z use. Galactic extinction is small, $E(B-V)=0.027
$ (Schlegel, Finkbeiner, Davis 1998), and hence neglected.
After excluding 411 stellar-like objects, a total of 35143 objects, whose
magnitudes are brighter than $i'=26.15$ (5$\sigma$) in
the $\sim590$~arcmin$^2$ area, are contained in the final catalog.

\subsection{Foreground/Background Subtraction by Photometric Redshifts}

Our aim is to map out structures associated with the known twin
clusters at $z\sim1.27$.
It is therefore required to remove unassociated galaxies in
the foreground and background as much as possible, to maximize
the contrast of the structures on the projected sky.

Since spectroscopic measurements for the $\sim 35,000$
galaxies are not practical, we exploit photo-z technique
as an observationally efficient method to largely subtract
the foreground/background populations.
We input the $VRi'z'$ magnitudes to the HYPERZ code (Bolzonella, Miralles,
Pell\'{o} 2000) to get estimated redshifts for all the individual
galaxies in our $i'$-band selected sample.
This code uses Bruzual \& Charlot's (1993)
stellar evolutionary code (GISSEL98) to build synthetic templates of
galaxies for eight star formation histories:
an instantaneous burst, a constant
star-forming system, and six exponentially decaying SFRs with
time-scales of from 1 to 30~Gyr. These models assume solar metallicity and
the Miller-Scalo IMF (Miller, Scalo 1979), and internal reddening is
considered using the Calzetti et al.\ (2000) model with $A_V$ varying
between 0 and 1.2 mag.

It is ideal to have near-infrared magnitudes as well as optical
magnitudes to get better estimates for photo-z's for galaxies
beyond redshift of unity,
since the prominent spectral break feature at 4000\AA\ for old galaxies
starts to range out from the optical passbands at $z>1$
(Kodama, Bell, Bower 1999; Connolly et al.\ 1997).
However, very limited FOVs of current near-infrared imagers on large
aperture telescopes (eg., $2'\times2'$ in the case of Subaru) make it
impractical to do near-infrared imaging of such a huge area of our
Suprime-Cam field ($\sim30'$).
Nevertheless, by obtaining $z'$-band magnitudes at the longest optical
wavelength, whose effective wavelength is about 9200\AA, we can
still deal with the galaxies
out to $z\sim1.3$ for photo-z estimates.
Our target supercluster is supposed to be located at $z\sim1.27$, hence
the $z'$-band can still just catch the 4000\AA\ break features.
In other words, this target is one of the most distant structures that can be
explored by photo-z's based on the optical imagings alone hence
with this huge field of view, except for the rest-UV-selected Lyman
break galaxies at exotically high redshifts such as $z>3$ (eg.,
Steidel et al.\ 1999).

By using the best estimate of photo-z of individual galaxies,
we select the only galaxies within the range of $1\leq z_{\rm phot}\leq1.35$
to isolate the plausible members associated to the supercluster at
$z\sim1.27$. The number of these photometrically selected candidates
for cluster members (hereafter photo-z selected candidates) is 2229.
This rather broad range of redshift is intended
to ensure that we include the bulk of the cluster population, allowing
for the intrinsic errors on photo-z's.
In fact, we find that 14 out of the 16 known true members of the twin
cluster, which are confirmed by spectroscopy (Stanford et al.\ 1997; Rosati et
al.\ 1999), are assigned photo-z's within the above range.
We therefore estimate that the completeness of the members of our method 
is greater than 80\%.
For blue cluster members, the estimated photo-z's tend to be smaller
than the true values. It is probably because our photo-z technique
misidentifies some blue cluster members as red galaxies at smaller
redshifts (Kodama et al.\ 1999).
For this reason, the distribution of $z_{\rm phot}-z_{\rm spec}$ is
likely to be asymmetric around zero. In fact, among the 14
spectroscopically confirmed galaxies, the standard deviation of 5
galaxies with $z_{\rm phot}>z_{\rm spec}$ is 0.05, while that of 9
galaxies with $z_{\rm phot}<z_{\rm spec}$ is 0.17.
Because of this, we take an asymmetric range for photometric members
around $z=1.27$ (ie., $1\leq z_{\rm phot}\leq1.35$).
The remaining 2 galaxies are
assigned significantly lower redshifts around $z\sim0.55$. Both of these
galaxies have foreground objects near to their images, and the photometry
of these member galaxies may be affected by these foreground objects.

%
%

\begin{figure}
\includegraphics[trim=1mm 40mm 1mm 40mm, clip, width=8cm]{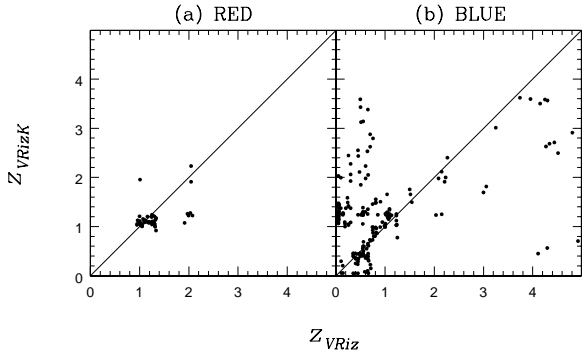}
\caption{Comparison between the photometric redshifts estimated with and
without the $K'$-band data ($Z_{VRizK}$ and $Z_{VRiz}$, respectively),
for the galaxies in the cl1, cl2 and gr3 down to $K'_{\rm AB}<23$
(see text for the data).
Panel (a) and (b) shows the red and blue galaxies, respectively,
separated at the threshold colour (see text).}
\label{fig1}
\end{figure}

To assess our photo-z's based on the optical data alone
($Z_{VRiz}$), we compare them to the photo-z's estimated WITH
the $K'$-band data ($Z_{VRizK}$), which are available for the selected
areas of our fields.
It is expected that the photo-z's at $z\sim1.27$ with the
$K'$-band have better accuracies since the $K'$-band neatly covers the
longwards of the rest-frame 4000\AA\, break at this redshift.
Here, we use the $K'$-band imaging data of the 3 regions
($2'\times2'$ each; centered on cl1, cl2 and gr3 in Figure~\ref{fig2})
taken by ourselves with a near-infrared camera CISCO on the Subaru telescope
The integration time of each image is about 30 min
and the limiting magnitude is about $K'_{\rm AB}=23$ ($5\sigma$).
The comparison is shown in Figure~\ref{fig1}.
The red and blue galaxies indicate the galaxies redder and bluer respectively
than a threshold colour, $(i'-z')=1.05-0.008z'$, which corresponds to the
typical colour of the local Sab galaxies (Fukugita, Shimasaku, Ichikawa
1995). The numbers of the red and the blue galaxies plotted in
Figure~\ref{fig1} are 48 and 200, respectively.
For the red galaxies, we have good match between $Z_{VRiz}$ and $Z_{VRizK}$,
suggesting the accuracy of our optical photo-z's are reliable
enough for these red populations.
In fact, the standard deviation, average, and median of $Z_{VRiz}-Z_{VRizK}$
for the red galaxies is $0.15$, $+0.06$, and $+0.04$, respectively, 
if we exclude six highly discrepant galaxies with $|Z_{VRiz}-Z_{VRizK}|>0.5$.
For the blue galaxies, however, we see a significant scatter, in particular,
we tend to underestimate photo-z's if we lack the $K'$-band.
Therefore, we must be careful in treating the blue galaxies.
In this paper, however, we rely basically on the red galaxies, and hence
this is not a big problem.

Given the broad redshift cut for photometric membership, the above
photo-z selected candidates should include significant fraction of field 
contamination which are not associated to the supercluster.
Later in \S~3, we discuss the level of field contamination in the
structures newly found in this study.
We stress, however, this photometric selection has already removed
the majority (32914/35143$\sim$95\%) of all the
foreground/background galaxies
while keeping the bulk ($\sim$80\%) of the cluster members.
This essential process has now made it possible for us to unveil the
large scale structure possibly associated to the known twin
cluster and spread
over the entire Suprime-Cam field, which are otherwise embedded in the
foreground/background galaxies and invisible.

\section{Large Scale Clumpy Structure}

%
%

\begin{figure}
\includegraphics[trim=1mm 5mm 1mm 5mm, clip, height=7cm]{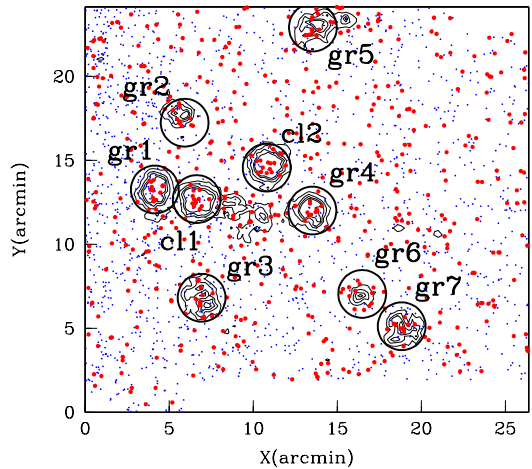}
\caption{The 2-D distribution of photo-z selected candidates on the sky.
North is up, and east is to the left. The large and small dots
indicate red and blue galaxies respectively. The contour levels are 2,
3, 4 and 5$\sigma$. See the text for details.
The radii of large circles are 0.7$h^{-1}_{70}$~Mpc at $z=1.27$.}
\label{fig2}
\end{figure}

Figure~\ref{fig2} shows the spatial distribution on the sky of the
photo-z selected candidates at $1\leq z_{\rm phot}\leq 1.35$.
The large and small dots indicate red and blue galaxies respectively
defined in \S2.2. The numbers of the red and blue galaxies are 575 and
1654, respectively. The positions of the galaxies, $X$ and $Y$, in the
Figure~\ref{fig2} are
expressed relative to the bottom-left corner of the Suprime-Cam image.
The contours indicate the local surface density of the
`red' photo-z selected candidates calculated from 10 nearest neighbors,
corresponding to 2, 3, 4 and 5$\sigma$.
Here $\sigma$ corresponds to the scatter (standard deviation) of the local
surface density in the case of the random distribution of 575 galaxies
over the 590~arcmin$^2$ field of view, ie., 0.97~arcmin$^{-2}$.
We do not include the blue galaxies here for the following reasons: (1) The
accuracy of photo-z is relatively poor for the blue galaxies
because of their intrinsically weak 4000\AA\ feature (\S2.2). (2)
The red galaxies trace clusters or groups of galaxies more neatly than the
blue galaxies due to the morphology(colour)-density relation (Dressler 1980,
Dressler et al.\ 1997). In fact, many authors have exploited this
technique and have successfully identified distant cluster candidates
in their wide-field data (e.g., Gladders, Yee 2000; Goto et al.\ 2002).
We note that the surface density of the blue galaxies is more than twice
larger in the area of $X\lesssim4.5$ (arcmin) compared to the other
area, while the surface density of the red galaxies are nearly the same.
This probably comes from the fact that the CCD chips were different between
these areas at the time of the observation; in the early phase, the
Suprime-Cam consisted of two types of CCDs, SITe (ST-002) and MIT/LL (CCID20).
The sensitivities of these 2 CCDs are different especially in the
$z'$-band, where the limiting magnitudes differ by $\sim0.3$ mag.
Hence the number of detected galaxies at $X\lesssim4.5$ with MIT/LL CCD's
are increased compared to the other area.
Since we count only red galaxies to define local surface density, however,
the contours in Figure~\ref{fig2} are not affected by this effect.

In Figure~\ref{fig2}, we mark by large circles the areas where the local
surface densities of the red galaxies are higher than 5$\sigma$ above
the mean density. The cl1 and cl2 indicate the 2
known high redshift clusters at $z\sim1.27$, RXJ~0848.9+4452 and
RXJ~0848.6+4453, respectively. We newly find 7 candidates
of galaxy clusters or groups, where the local surface density is as large
as those of the known clusters (gr1-gr7).

%
%

\begin{figure}
\includegraphics[trim=1mm 0mm 1mm 0mm, clip, width=7cm]{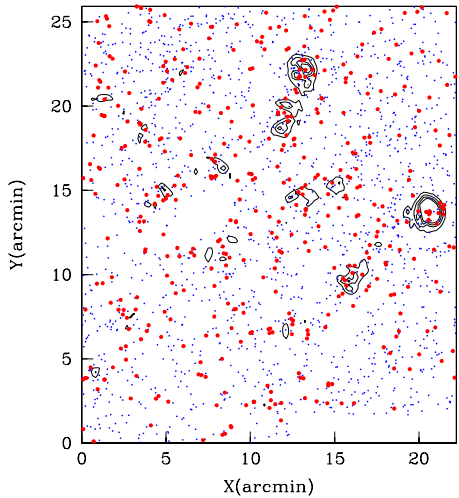}
\caption{The 2-D distribution of photo-z selected galaxies on the sky in the
Subaru Deep Field. North is up, and east is to the left. The symbols and the
contour levels are the same as in Figure~\ref{fig2}. See the text for details.}
\label{fig3}
\end{figure}

As we noted in \S~2.2, our photo-z selected member candidates should
include significant amount of field contamination
due to the generous criterion for photometric membership 
($1\leq z_{\rm phot}\leq1.35$).
However, the contrast of our new cluster/group candidates
against the general field is always statistically significant.
The number density of photo-z selected candidates of red galaxies
in the low density ($<0\sigma$) regions is
$2.5\pm0.2~h^2_{70}$~Mpc$^{-2}$, while those of the
high density regions within the circles of 0.7$h^{-1}_{70}$~Mpc radii
centered on our cluster/group candidates including the known twin clusters
range from $6.3\pm2.0~h^2_{70}$~Mpc$^{-2}$ to $11.2\pm2.7~h^2_{70}$~Mpc$^{-2}$.
To verify the significance of the structures, we present the typical galaxy
distribution at $z\sim1.3$ in the blank field for comparison using the
Subaru Deep Field data (SDF; e.g., Furusawa 2002; Kashikawa et al.\
2003; Ouchi et al.\ 2003) in Figure~\ref{fig3}. Here we have applied
the photo-z using the same passbands ($VRi'z'$) and have extracted
only galaxies that fall within $1\leq z_{\rm phot}\leq 1.35$, the same
criteria that we used for the Lynx cluster field. The contours
indicate the surface number density of the red galaxies within this
redshift slice and the contour levels are the same as in
Figure~\ref{fig2}. As shown, only the two areas with $>$5$\sigma$
peaks are found in the SDF, while 9 such regions are identified in the
Lynx field. It is therefore likely that most of the structures that we
found in the Lynx field are real.
To further strengthen the reality of the large scale structures
associated to the twin cluster, we present below the colour-magnitude diagram,
radial profile, and the estimate for cluster/group richness for each clump.

%
%

\begin{figure}
\includegraphics[trim=1mm 0mm 1mm 0mm, clip, height=7.5cm]{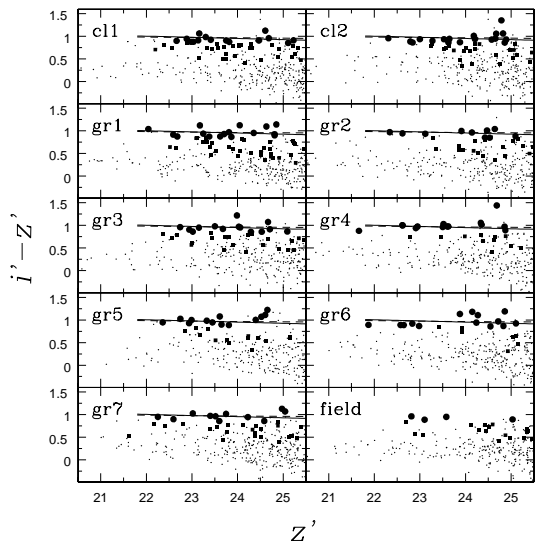}
\caption{The $i'-z'$ versus $z'$ diagrams of the galaxies in the known clusters
(cl1 and cl2) and our new cluster/group candidates (gr1-gr7).
See the text for details.}
\label{fig4}
\end{figure}

Figure~\ref{fig4} shows the $i'-z'$ versus $z'$ diagram for each
cluster/group candidate including the two known clusters.
Galaxies within a circle of 0.7$h^{-1}_{70}$~Mpc radius are plotted.
The filled circles and the filled squares indicate the photo-z
selected candidates of red and blue galaxies, respectively. The small dots
show the other galaxies. For comparison, we also show the galaxies
randomly selected from the low density ($<0\sigma$) regions in the
bottom right panel of Figure~\ref{fig4}.
The number of galaxies shown in this panel is the expected number of
galaxies within a circle of 0.7$h^{-1}_{70}$~Mpc radius calculated from the
surface density of galaxies in the Lynx field.
The solid line in each panel shows the red sequence of the
cluster RDCS~1252 at $z=1.24$ derived by Blakeslee et al.\ (2003).
We applied an appropriate colour-term correction from HST/ACS filters to
Suprime-Cam ones. The dashed line in each panel indicates a predicted
colour-magnitude relation at $z=1.27$ for passively
evolving galaxies formed at $z_{\rm F}=4.5$ (Kodama, Arimoto 1997),
although they are almost identical to the empirically fitted
lines (solid lines).
Not only in the two known clusters (cl1 and cl2) but also in our
new candidates (gr1-gr7), can we find many red galaxies consistent
with the passively evolving galaxies
which comprise the red sequences in these diagrams.

%
%

\begin{figure}
\includegraphics[trim=1mm 15mm 1mm 15mm, clip, width=8cm]{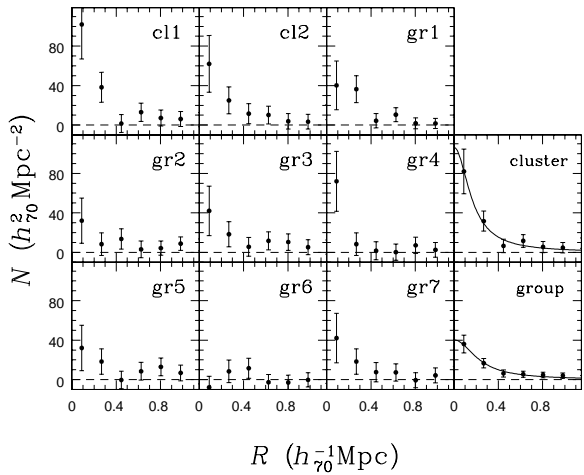}
\caption{The surface number density of all galaxies as a function of the
distance from the cluster center. Cluster and group indicate the combined
profiles of the two known clusters and those of our new
candidates, respectively, and
the best-fitted King's law profiles are also plotted by the solid lines.}
\label{fig5}
\end{figure}

We also show the radial profile of each cluster and cluster/group
candidate in Figure~\ref{fig5}.
Only those galaxies brighter than $i'=26$ are counted because of increasing
incompleteness toward fainter magnitudes.
Here we subtract the remaining field contamination using the SDF data
(see above) from each bin after normalizing counts of field galaxies
with referring areas. The averaged surface
number densities of all galaxies and the red galaxies are
$7.8\pm7.1$ and $0.39\pm0.65~h^2_{70}$~Mpc$^{-2}$,
respectively, to the same magnitude limit of $i'=26$.
Given the large field of views, the field-to-field variation of the
galaxy densities should be averaged over, if present, and hence the
accuracy of statistical field subtraction should not be largely affected
by this effect.
We see a pronounced increase in galaxy density towards the center of most
of our cluster/group candidates.
The only exception is gr6 that shows no excess of galaxy density at the
center. This may be due to the misidentification of the center of the group,
or intrinsically non-axisymmetric structure.
We combine the profiles of known clusters and those of our new
candidates (groups) separately,
and find that both of these combined profiles can be fitted very well
with the King's law profile, $N(r)=N_0(1+r^2/R_c^2)^{-1}$ (King 1966, 1972),
with $R_{\rm c}=0.17\pm0.02$~$h_{70}^{-1}$~Mpc (cluster) and $R_{\rm c}=0.23
\pm0.03$~$h_{70}^{-1}$~Mpc (group), which is consistent with estimations for
local clusters (Bahcall 1975).

%
%

\begin{table*}
\caption{Properties of the clusters and the cluster/group candidates.
Units of right ascension are hours, minutes and seconds, and
units of declination are degrees, arcminutes and arcseconds.
The local surface densities ($\Sigma,$ $\Sigma_{\rm red}$) given in the table
indicate the ones after the remaining field contaminations have been
subtracted (see \S3 for detail). In the table, $H_0=70$~km~s$^{-1}$~Mpc$^{-1}$,
$\Omega_0=0.3$ and $\Lambda_0=0.7$ are assumed.}
\label{table2}
\begin{tabular}{lcccccccc}
\hline
 & R.A. & Decl. & $\Sigma$ & $\Sigma_{\rm red}$ & $N_{0.5}$ & $R$ &
 $F_X$(0.5-2.0~keV)$^a$ & $L_X$(0.5-2.0~keV)$^a$ \\
 & (J2000.0) & (J2000.0) & (Mpc$^{-2}$) & (Mpc$^{-2}$) & & &
 ($10^{-15}$~ergs~s$^{-1}$~cm$^{-2}$) & ($10^{44}$~ergs~s$^{-1}$) \\
\hline
  cl1     & 08 48 57.3 & +44 52 02.8 & $20.2\pm8.3$  & $9.0\pm2.5$
          & $23.6\pm7.6$   & 1-2   & $3.8\pm0.8^b$  & $0.36\pm0.08$ \\
  cl2     & 08 48 34.3 & +44 53 52.3 & $17.1\pm8.1$  & $10.2\pm2.7$
          & $19.6\pm7.3$   & 1     & $4.7\pm0.5^b$  & $0.44\pm0.05$ \\
  gr1$^c$ & 08 49 11.7 & +44 52 37.0 & ($11.5\pm6.9$) & ($9.6\pm2.6$)
          & ($15.4\pm7.0$) & (0-1) & $1.0\pm0.4^d$  & $<0.22^e$ \\
  gr2     & 08 49 01.3 & +44 56 33.0 & $9.6\pm7.9$   & $5.8\pm2.1$
          & $6.6\pm6.4$    & 0     & $0.0\pm0.4^d$  & $<0.12^e$ \\
  gr3     & 08 48 57.7 & +44 46 09.5 & $13.4\pm8.0$ & $8.3\pm2.4$
          & $10.5\pm6.9$   & 0     & $-0.2\pm0.4^d$ & $<0.10^e$ \\
  gr4     & 08 48 18.7 & +44 51 18.7 & $7.1\pm7.8$  & $7.7\pm2.3$
          & $7.1\pm5.2$    & 0     & $0.4\pm0.4^d$  & $<0.16^e$ \\
  gr5     & 08 48 18.5 & +45 02 13.2 & $9.5\pm7.9$  & $8.0\pm2.4$
          & $9.6\pm6.6$    & 0     & $0.0\pm0.3^d$  & $<0.10^e$ \\
  gr6     & 08 48 02.0 & +44 46 23.1 & $4.0\pm7.6$  & $7.1\pm2.3$
          & $4.6\pm5.4$    & 0     & $\cdots$     & $\cdots$ \\
  gr7     & 08 47 48.7 & +44 44 27.6 & $12.1\pm8.0$ & $5.2\pm2.0$
          & $14.9\pm6.8$   & 0-1   & $\cdots$     & $\cdots$ \\
\hline
\end{tabular}
\begin{flushleft}
\hspace{1cm}$^a$ $r=35''$ aperture \\
\hspace{1cm}$^b$ quoted from Stanford et al.\ (2001) \\
\hspace{1cm}$^c$ The tabulated numbers for gr1 may be significantly
affected by the contamination of the blue galaxies in the
$X\lesssim4.5$~(arcmin) region, and hence \\
\hspace{1.1cm}unreliable. \\
\hspace{1cm}$^d$ The flux (corrected for the Galactic absorption) and
luminosity are given in the rest-frame 0.5--2 keV band. They are
converted from the 0.5--2 keV \\
\hspace{1.1cm}count rate in the observed frame by assuming an
optically-thin thermal plasma model at $z=1.27$ with a temperature of
6 keV and elemental \\
\hspace{1.1cm}abundances of 0.3 solar. \\
\hspace{1cm}$^e$ 3$\sigma$ upper limits
\end{flushleft}
\end{table*}

Table~\ref{table2} summarizes the properties of the known clusters and
the cluster/group candidates.
The J2000.0 coordinates of the each cluster/group candidate are
given in columns 2 and 3.
Columns 4 and 5 show the surface number density of all the galaxies and the
red galaxies only, respectively, within a circle of
$0.7h^{-1}_{70}$~Mpc radius. We also only use galaxies at $i'<26$.
The remaining field contamination has been subtracted using the
SDF data (see above), and the errors indicate the
Poisson statistics based on the number of cluster galaxies and that of
the subtracted field galaxies.
In column 6, we show the estimated richness of each cluster/group candidate,
using the $N_{0.5}$ indicator introduced by Hill \& Lilly (1991).
This indicates the number of galaxies within a 0.5~Mpc radius
from the cluster center and within the magnitude range between $m_1$ and
$m_1+3$ measured in the $R$ band at $z\sim0.5$.
Since the $R$ band at $z\sim0.5$ roughly
corresponds to the $z'$-band at $z\sim1.27$,
we can make a direct comparison with their measurements.
The cluster richness estimated based on $N_{0.5}$ is 1-2 for cl1 and
cl2, while 0-1 for gr1-gr7 (See Table~4 of Hill \& Lilly 1991).
Therefore we find that the candidates
we newly find are poorer systems than the two known clusters.
This is also suggested from deep X-ray data around the
Lynx field taken with Chandra (Stanford et al.\ 2001; Stern et al.\ 2002).
Using the Chandra image in the 0.5--2 keV band, we measure the flux of
excess diffuse emission in the circular region around each candidate
with a radius of 35$''$. The background (the unresolved X-ray background
and non X-ray background) is estimated from a surrounding annular region.
The result is shown in column~8 of Table~\ref{table2}.
Note that gr6 and gr7 are outside of the Chandra field. 
We cannot detect any significant X-ray excess at more
than the $3\sigma$ level for our cluster/group candidates except for 
the two known clusters.
It should, however, be noted that the $3\sigma$ upper limit 
luminosity of each region (see column~9 of Table~\ref{table2}) 
is comparable to a typical luminosity of nearby groups of galaxies. 
Thus, the non-detection in the Chandra data does not necessarily imply 
that the regions we find have unusually low X-ray luminosities
compared with their optical richnesses.
We argue therefore that our `Suprime-Cam imaging + photo-z'
method is a very powerful technique in finding clusters or groups
in the distant Universe even beyond $z>1$ out to $z\sim1.3$.
We note that this is practically the highest redshift
structures of `normal' galaxies that can be traced by optical images
with our scheme (bracketing the 4000\AA-break region in the rest frame).

\section{Summary}

We have presented the results of deep panoramic imaging of the Lynx 
supercluster field
at $z\sim1.27$ taken with the Suprime-Cam on Subaru Telescope.
Our multicolour image covers an area of 
$26'.4\times24'.1$ ($13.3\times12.1h_{70}^{-1}
$ Mpc), allowing us for the first time to investigate large scale
structure spreading around the known cluster regions at this high redshift.
By applying the photometric-redshift technique, we have mapped out the spatial
distribution of galaxies near the redshift of the supercluster 
down to $\sim M^\ast+2.5$.

We have newly discovered seven cluster/group candidates.
These candidates show red colour-magnitude sequences and centrally
concentrated  profiles, similar to those of the two known clusters in
this field, indicating that they are likely to be real clusters/groups
of galaxies around the Lynx supercluster,
comprising a large-scale structure over $\sim$13~$h^{-1}_{70}$Mpc scales.
The clumpiness of the candidates suggests that they are in early dynamical 
stages, being in the process of assembly to a massive cluster. 

The next key step will be to confirm with deep spectroscopic observations
that the cluster/group candidates we discovered in this study are physically
associated with one another and forming 
a real large-scale structure at $z\sim 1.27$.
If confirmed, this will be the highest redshift structure of
`normal' galaxies ever identified, which will have a large impact on
structure formation theories and will bring us invaluable
information on the formation and evolution of clusters of galaxies
at their early stages.

\section*{Acknowledgments}

We thank R.\ Ellis, R.\ Bower, M.\ Balogh, T.\ Yamada and W.\ Couch
for useful discussion. FN acknowledge the financial support from UK
PPARC. FN, TK and MO acknowledge support from Japan
Society for the Promotion of Science through its research fellowships
for young scientists. SO and TK acknowledge the Daiwa-Adrian Prize
2001 given by The Daiwa Anglo-Japanese Foundation. This work was
financially supported in part by a Grant-in-Aid for the Scientific
Research (No.15740126) by the Japanese Ministry of Education, Culture,
Sports and Science.

\label{lastpage}

\end{document}